\documentclass[fleqn]{llncs}
 \pdfoutput=1 
\usepackage{url}
\usepackage{color}
\usepackage{amsmath}

\usepackage{algorithm}
\usepackage[noend]{algpseudocode}

\newcommand{\set}[1]{\left\{
      \begin{array}{l}#1\end{array}
      \right\}}
\newcommand{\sset}[2]{\left\{~#1  \left|
      \begin{array}{l}#2\end{array}
    \right.     \right\}}

\renewcommand{\phi}{\varphi}

\definecolor{cadmiumgreen}{rgb}{0.0, 0.42, 0.24}
\newcommand{\ignore}[1]{}
\newcommand{\bee}{\textsf{BEE}}

\title{A SAT Encoding for the $n$-Fractions Problem}
\author{Michael Codish}
\institute{
  Department of Computer Science,
  Ben-Gurion University of the Negev, Israel
}

\begin{document}

\maketitle

\begin{abstract}
  This note describes a SAT encoding for the $n$-fractions puzzle
  which is problem \texttt{041} of the CSPLib. Using a
  SAT solver we obtain a solution for two of the six remaining open
  instances of this problem.
\end{abstract}

\section{Introduction}

The $n$-fractions puzzle \cite{csplib:prob041} is problem \texttt{041} of
the CSPLib. The original puzzle is specified as follows: find
nine distinct non-zero digits, $\{A,B,C,D,E,F,G,H,I\}$, that satisfy
\[ 
  \frac{A}{BC} + \frac{D}{EF} + \frac{G}{HI}  = 1
\]
where $BC$ is shorthand for $10B +C$, $EF$ for $10E +F$, and
$H$ for $10H + I$.
A simple generalization is as follows: find $3n$ nonzero
digits, $x_i,y_i,z_i$ ($1\leq i\leq n$), satisfying
\begin{equation}\label{eq:problem_specification}
    \sum_{i=1}^n \frac{x_i}{y_iz_i} = 1
  \end{equation}
where $y_iz_i$ is shorthand for $10y_i + z_i$ and the number of
occurrences of each digit in $\{1,\ldots,9\}$ is between 1 and
$\lceil n/3\rceil$. 
An interesting problem is to find the greatest $n$ such that at
least one solution exists.
Since each fraction is at least $1/99$, this family of problems has
solutions for at most $n\leq 99$. Malapert and Provillard prove in a
recent paper~\cite{malapert2017puzzle} that the puzzle has no solution for $n\geq
45$. 

Two models are described in the literature~(see \cite{malapert2017puzzle}) to solve the
$n$-fractions puzzle. 
The \emph{division model} handles
Equation~\eqref{eq:problem_specification} with floating point
arithmetic. This approach returns invalid solutions because of
rounding errors. 
The \emph{product model} only needs integer arithmetic because
Equation~\eqref{eq:problem_specification} is reformulated as follows:
\begin{equation}
  \label{eq:product_model}
  \sum_{i=1}^n \left(x_i\prod_{k\neq i} y_kz_k\right) = \prod_{i=1}^n y_kz_k
\end{equation}
The main problem with the product model is that the number of bits
required to represent the products grows exponentially with the size
of $n$. For example, the multiplication term on the right side of
Equation~\eqref{eq:product_model} overflows a 32-bit integer for
$n=6$.

Malapert and Provillard~\cite{malapert2017puzzle} propose an \emph{integer
  factorization model} and demonstrate that applying this model they
can find solutions for all of the instances with $n<45$ except for
six: where $n\in\{36,39,41,42,43,44\}$.
Their approach comprises two basic ideas:
The first idea is to solve the following constraint instead of
  that expressed as Equation~\eqref{eq:problem_specification}:
\begin{equation}
  \label{eq:specification_with_lcm}
  \sum_{i=1}^n x_i\times \frac{L}{y_iz_i} = L
\end{equation}
where $L$ is the lowest common multiple of the integers
$\sset{y_iz_i}{1\leq i\leq n}$. In this formalization, each of the
terms, $\frac{L}{y_iz_i}$ on the left side of
Equation~\eqref{eq:specification_with_lcm} is an integer. In theory,
the products in Equation~\eqref{eq:specification_with_lcm} still grow
exponentially. In practice, based on this formulation, it is possible
to solve large $n$-fractions puzzles.
The second idea is to represent the integer variables in
  Equation~\eqref{eq:specification_with_lcm} in terms of their prime
  factorizations.

In this note we describe a simple \emph{LCM model} for the
$n$-fractions problem. The approach is based on
Equation~\eqref{eq:specification_with_lcm}. We encode the constraints
of this model to SAT using a standard binary representation for
integers. Our approach is able to solve two of the instances left open
in the paper by Malapert and Provillard~\cite{malapert2017puzzle}. These are the
36-fraction puzzle and the 39-fraction puzzle.

\section{The LCM Constraint Model}

In this section we describe a simple \emph{LCM model} for the
$n$-fractions problem in terms of finite integer constraints.  These
are then compiled to CNF using the finite-domain constraint compiler
\bee~\cite{Metodijair2013} which compiles constraints to CNF. The
(conjunctions of) constraints in our model (in \bee\ syntax) are
detailed below as framed text.

\subsection{Domain and Counting Constraints}
\label{section:domain}
For $1\leq i\leq n$, the variables $x_i,y_i,z_i$ take integer values
in the domain $\{1,\ldots,9\}$. The number of occurrences of each
digit is constrained to be between 1 and $\lceil{n/3}\rceil$.
The variables $y_iz_i = 10\times y_i+z_i$ take integer values in the
domain $\{11,\ldots,99\}$. 

In \bee\ an integer variable $\mathtt{x}$ is declared to be in unary
or binary representation, $\mathtt{new\_int(x,lb,ub)}$ or
$\mathtt{new\_binary(x,lb,ub)}$, where $\mathtt{lb}$ and $\mathtt{ub}$
are lower and upper bounds. 

The variables $x_i,y_i,z_i$ and $y_iz_i$ are represented in unary
representation.  The variables $x_i$ and $y_iz_i$ are represented also
through channelling to their binary representation. This is because
the counting constraints (on the digits) are best encoded to CNF using
the unary representation while the arithmetic constraints described in
Sections~\ref{section:lcm_constraints}
and~\ref{section:puzzle_constraints} are best encoded to CNF using the
binary representation.
In the constraint model, detailed as Figure~\ref{modelpart1}, we denote
the digits $[x_1,\ldots,x_n, y_1,\ldots,y_n, z_1,\ldots,z_n,]$ by
$[dig_1,\ldots,dig_{3n}]$ and then the (Boolean) variables $dig_{i,j}$
denote that $dig_i$ takes value $j$ and the (integer) variables $s_j$
denote the number of occurrences of the value $j$ among
$[dig_1,\ldots,dig_{3n}]$ (for $1\leq i\leq 3n, 1\leq j\leq 9$).

\begin{figure}[t]
  \centering
  \fbox{\[ \begin{array}{l} 
  \displaystyle\bigwedge_{i=1}^n 
  \set{new\_int(x_i,1,9), new\_int(y_i,1,9), new\_int(z_i,1,9), new\_int(y_iz_i,11,99), \\
       channel\_int2binary(x_i), channel\_int2binary(y_iz_i),\\
       int\_array\_lin\_eq([10,1],[y_i,z_i],y_iz_i)
      } \\
   \displaystyle\bigwedge_{i=1}^{3n} \bigwedge_{j=1}^9  
   \set{int\_eq\_reif(dig_i,j,dig_{i,j})
       } \\
   \displaystyle\bigwedge_{j=1}^9 
   \set{new\_int(s_j,1,\lceil n/3\rceil),
        bool\_array\_sum\_eq([dig_{1,j},\ldots dig_{3n,j}],s_j)
       }
   \end{array}
\]}
  
  \caption{The \bee\ model: part~1.}\label{modelpart1}
\end{figure}

\subsection{Symmetry Breaking and Redundant Constraints}
\label{section:symmetry}

We add the symmetry breaking constraints and a redundant constraint
proposed by Frisch~\cite{frisch2004symmetry}
\begin{equation}\label{eq:symmetry}
  (y_i,z_i,x_i)\leq_{lex}(y_{i+1},z_{i+1},x_{i+1})~~ 1\leq i<n
\end{equation}
\begin{equation}
  \min_{1\leq i\leq n}y_iz_i \leq \sum_{i=1}^nx_i\leq \max_{1\leq i\leq n}y_iz_i
\end{equation}
For the \bee\ syntax see Figure~\ref{modelpart2}.

\begin{figure}[b]
  \centering
  \fbox{\[ \begin{array}{l} 
    \displaystyle\bigwedge_{i=1}^{n-1}
    \set{int\_arrays\_lex([y_i,z_i,x_i],[y_{i+1},z_{i+1},x_{i+1}])
        } \\
    \bigwedge\set{new\_int(r,n,9n),  int\_array\_sum\_eq([x_1,\ldots,x_9],r), \\
               new\_int(min,11,99), int\_array\_min([y_1z_1,\ldots y_nz_n],min),\\
               new\_int(max,11,99), int\_array\_max([y_1z_1,\ldots y_nz_n],max),\\
               int\_leq(min,r),   int\_leq(r,max)
        }
      \end{array}
  \] }
  
  \caption{The \bee\ model part 2.}\label{modelpart2}
  
\end{figure}

\subsection{LCM Constraints}
\label{section:lcm_constraints}

The least common multiple, $L$ of a set of positive integers $S$ is
the smallest positive integer that is divisible by each of the
integers in $S$. In the context of
Equation~\eqref{eq:specification_with_lcm}, it is sufficient if $L$ is
any common multiple.

We introduce integer variables $L$ and $\{d_1,\ldots,d_n\}$. The
variable $L$ takes values in the domain $\{1,\ldots \mathit{maxL}\}$
where $\mathit{maxL}$ is a parameter of the encoding. The variables
$d_i$ take values in the domain $\lceil \mathit{maxL}/11 \rceil$.  The
following constraint states that $L$ is divided by each of the
numbers $y_iz_i$. This constraint also ``determines'' the variables
$d_i$, or more precisely, the relation between the variables
$y_iz_i,L$ and $d_i$.
\begin{equation}
  \label{eq:lcm}
  \bigwedge_{i=1}^n y_iz_i \times d_i = L
\end{equation}

For an optimization, we observe that often many of the values in the
sequence $y_1z_1,\ldots y_nz_n$ are repeated (see
Table~\ref{tab:solutions}). Moreover, because of the specific symmetry
break of Equation~\eqref{eq:symmetry}, repeated values $y_iz_i$ occur
consecutively in this sequence.
Instead of encoding the LCM constraints using Equation~\eqref{eq:lcm},
we encode them with the following constraints
\begin{equation}
  \bigwedge_{i=1}^{n-1} \mathtt{if~}(y_iz_i=y_{i+1}z_{i+1})
  \mathtt{~then~} (d_i=d_{i+1}) \mathtt{~else~}(y_iz_i \times d_i = L)
\end{equation}

In Figure~\ref{modelpart3}, the variables $[\ell_1,\ldots,\ell_n]$ are
such that $y_iz_i\times d_i=\ell_i$. If we constrain all of the $\ell_i$ to
equal $\ell_1$ then $\ell_1$ is a common multiplier of the divisors
($y_i,z_i$). Instead we only constrain $\ell_i=\ell_1$ where the
divisor $y_iz_i$ occurs first (not repeated) in the sequence of
divisors.

\begin{figure}[t]
  \centering
  \fbox{\[ \begin{array}{l} 
    \displaystyle\bigwedge_{i=1}^{n}
    \set{new\_binary(\ell_i,1,maxL), new\_binary(d_i,1,\lceil maxL/ 11 \rceil), \\
         binary\_times(y_iz_i,d_i,\ell_i)
        } \\
    \displaystyle\bigwedge_{i=1}^{n-1}
    \set{binary\_eq\_reif(y_iz_i,y_{i+1}z_{i+1},a_i),
         binary\_eq\_reif(d_i,d_{i+1},b_i),\\
         binary\_eq\_reif(\ell_{i+1},\ell_1,c_i),
         bool\_array\_or([-a_i,b_i]),
         bool\_array\_or([a_i,c_i])
        }
      \end{array}
      \]}
  \caption{The \bee\ model part 3.}
  \label{modelpart3}
\end{figure}

\subsection{The Puzzle Constraint}
\label{section:puzzle_constraints}

\begin{figure}[b]
  \centering
  \fbox{\[ \begin{array}{l} 

    \displaystyle\bigwedge_{i=1}^{n}
    \set{new\_binary(t_i,1,(9/11)\times maxL), binary\_times(x_i,d_i,t_i)
        }\\
    \bigwedge\set{binary\_array\_sum\_eq([t_1,\ldots,t_n],\ell_1)}  \\[7mm]

  \end{array}\]}
  \caption{The \bee\ model part 4.}
  \label{modelpart4}
\end{figure}

Equation~\eqref{eq:specification_with_lcm} is
modeled by the following constraint expressed in terms of the
variables $d_i$ introduced in the model as described in
Section~\ref{section:lcm_constraints}.
We encode Equation~\eqref{eq:problem_specification} as 
\begin{equation}
 \sum_{i=1}^n x_i\times d_i = L
\end{equation}
For the \bee\ syntax see Figure~\ref{modelpart4}.

\section{Experimental Results}

\begin{table}
  \centering
\begin{tabular}{|@{~~}rr@{~~}|@{~~}r@{~~}r@{~~}r@{~~}r@{~~}|}
\hline
n & maxL\hspace{-1mm} & \bee & \# cl~~ & \# var & sat~ \\
\hline

 3 &  300 & 0.05 &    10954 &   1663 &      0.11 \\
 4 &  100 & 0.07 &    14171 &   2054 &      0.03 \\
 5 &  100 & 0.09 &    18231 &   2596 &      0.04 \\
 6 &  100 & 0.16 &    22370 &   3122 &      0.09 \\
 7 &  100 & 0.15 &    27330 &   3788 &      0.13 \\
 8 &  100 & 0.15 &    31937 &   4341 &      0.21 \\
 9 &  100 & 0.28 &    36661 &   4915 &      0.10 \\
10 &  100 & 0.27 &    42207 &   5526 &      0.24 \\
11 &  100 & 0.29 &    47414 &   6143 &      0.28 \\
12 &  100 & 0.33 &    52313 &   6644 &      0.57 \\
13 &  100 & 0.25 &    58444 &   7373 &      0.30 \\
14 &  100 & 0.49 &    63490 &   7914 &      0.79 \\
15 &  120 & 0.35 &    71762 &   9466 &      6.95 \\
16 &  100 & 0.52 &    74966 &   9161 &      2.15 \\
17 &  100 & 0.45 &    79991 &   9759 &      1.79 \\
18 &  300 & 0.47 &    90836 &  11952 &      7.03 \\
19 &  100 & 0.54 &    91856 &  10988 &      6.03 \\
20 &  300 & 0.71 &   102790 &  13372 &     16.61 \\
21 &  300 & 0.80 &   108010 &  14060 &     28.08 \\
22 &  300 & 0.83 &   115090 &  14793 &    202.55 \\
23 &  300 & 0.57 &   120344 &  15506 &    257.02 \\
24 &  300 & 0.91 &   125787 &  16131 &     14.05 \\
25 &  300 & 1.06 &   132945 &  16977 &    374.90 \\
26 &  300 & 0.77 &   138824 &  17604 &    382.66 \\
27 &  400 & 1.12 &   147838 &  19311 &     16.70 \\
28 &  300 & 1.14 &   151870 &  19077 &    769.62 \\
29 &  400 & 1.23 &   161155 &  20856 &    951.97 \\
30 &  500 & 1.16 &   166856 &  21755 &    162.36 \\
31 &  500 & 1.31 &   174467 &  22638 &    253.78 \\
32 &  500 & 0.87 &   179809 &  23317 &    983.09 \\
33 & 1900 & 1.70 &   200607 &  28633 &   8427.08 \\
34 &  500 & 1.53 &   192702 &  24875 &   4690.07 \\
35 & 2400 & 1.91 &   217837 &  31579 &   6.11 hr \\
36 & 2400 & 1.16 &   223404 &  32427 &  37.99 hr \\
37 & 2400 & 1.04 &   185947 &  31670 &  67.77 hr \\
38 & 2400 & 1.99 &   237793 &  34322 &  66.58 hr \\
39 & 8400 & 4.86 &   326435 &  72219 & 102.20 hr \\

\hline
\end{tabular}
  \caption{Solving $n$-fractions with \bee}
  \label{tab:bee}
\end{table}

\begin{table} \tiny
  \centering

\begin{tabular}{|r|r|l|}
\hline
3  & 204& 9 12  5 34  7 68  \\
4  & 72 & 9 18  4 24  5 36  7 36  \\
4  & 54 & 3 18  6 18  9 27  9 54  \\
5  & 54 & 3 18  6 18  9 27  2 54  7 54  \\
6  & 72 & 4 18  5 18  4 36  5 36  9 72  9 72  \\
7  & 54 & 3 18  9 18  1 27  2 27  3 54  3 54  6 54  \\
8  & 68 & 2 17  2 17  5 34  5 34  9 34  2 68  5 68  7 68 \\ 
9  & 78 & 1 13  1 26  5 26  5 26  4 39  9 39  4 78  4 78  5 78  \\
10 & 56 & 1 14  1 14  3 28  3 28  3 28  3 28  4 56  4 56  7 56  9 56  \\
11 & 76 & 1 19  2 19  2 19  4 38  4 38  5 38  9 38  2 76  2 76  4 76  4 76  \\
12 & 78 & 1 26  1 26  4 26  4 26  4 39  4 39  5 39  5 39  1 78  1 78  5 78  5 78  \\
13 & 76 & 1 19  2 19  2 19  2 19  2 38  2 38  4 38  4 38  3 76  4 76  4 76  5 76  8 76  \\
14 & 78 & 1 26  1 26  1 26  4 26  2 39  4 39  4 39  4 39  4 39  1 78  5 78  5 78  5 78  5 78  \\
15 & 156 & 1 39  1 39  1 39  4 39  4 39  4 52  4 52  4 52  6 52  6 52  1 78  1 78  6 78  6 78  6 78  \\
16 & 78  & 1 26  1 26  1 26  1 26  3 26  1 39  1 39  2 39  4 39  5 39  4 78  4 78  4 78  5 78  5 78  
\\ & & 9 78 \\ 
17 & 96  & 1 32  1 32  1 32  1 32  2 32  1 48  1 48  3 48  5 48  5 48  5 48  5 96  5 96  7 96  7 96  
\\ & &7 96  7 96 \\ 
18 & 288 & 1 18  3 18  1 48  1 48  3 48  5 48  1 72  5 72  5 72  5 72  5 72  5 72  3 96  3 96  3 96  
\\ & &3 96  4 96  4 96 \\ 
19 & 96  & 1 32  1 32  2 32  3 32  3 32  1 48  1 48  1 48  1 48  1 48  5 48  5 48  2 96  5 96  5 96  
\\ & &5 96  5 96  7 96  7 96  \\
20 & 288 & 3 36  1 48  1 48  3 48  3 48  3 48  3 48  3 48  1 72  1 72  5 72  5 72  5 72  5 72  5 72  
\\ & &1 96  1 96  1 96  6 96  9 96  \\
21 & 288 & 1 18  1 48  1 48  3 48  3 48  3 48  3 48  5 72  5 72  5 72  5 72  5 72  5 72  5 72  1 96  
\\ & &1 96  1 96  3 96  3 96  3 96  4 96  \\
22 & 288 & 1 48  1 48  1 48  1 48  1 48  4 48  4 72  5 72  5 72  5 72  5 72  5 72  5 72  8 72  1 96  
\\ & &3 96  3 96  3 96  3 96  3 96  3 96  3 96  \\
23 & 288 & 1 18  1 24  1 48  1 48  1 48  2 48  3 48  3 48  3 48  3 72  3 72  3 72  3 72  3 72  5 72  
\\ & &1 96  1 96  5 96  5 96  5 96  5 96  5 96  5 96  \\
24 & 288 & 1 48  1 48  1 48  1 48  1 48  1 48  3 48  3 48  1 72  5 72  5 72  5 72  5 72  5 72  5 72  
\\ & &5 72  1 96  3 96  3 96  3 96  3 96  3 96  3 96  5 96  \\
25 & 288 & 1 48  1 48  1 48  1 48  1 48  5 48  5 48  5 48  1 72  3 72  3 72  3 72  3 72  4 72  5 72  
\\ & &5 72  1 96  1 96  1 96  2 96  3 96  3 96  3 96  3 96  3 96  \\
26 & 288 & 1 48  1 48  1 48  1 48  1 48  1 48  1 48  3 48  5 48  3 72  3 72  3 72  3 72  3 72  3 72  
\\ & &3 72  3 72  1 96  1 96  2 96  5 96  5 96  5 96  5 96  5 96  5 96  \\
27 & 380 & 1 38  1 38  1 38  1 38  2 38  2 38  2 38  4 38  4 38  1 76  1 76  1 76  1 76  2 76  2 76  
\\ & &4 76  4 76  4 76  1 95  2 95  2 95  2 95  2 95  4 95  4 95  4 95  4 95  \\
28 & 288 & 1 48  1 48  1 48  1 48  1 48  1 48  1 48  1 48  1 48  1 48  3 72  3 72  3 72  3 72  3 72  
\\ & &3 72  3 72  3 72  3 72  2 96  3 96  5 96  5 96  5 96  5 96  5 96  5 96  5 96 \\ 
29 & 380 & 1 38  1 38  1 38  1 38  2 38  2 38  2 38  2 38  4 38  1 76  1 76  1 76  1 76  2 76  2 76  
\\ & &4 76  4 76  4 76  4 76  1 95  1 95  2 95  2 95  2 95  2 95  3 95  4 95  4 95  4 95 \\ 
30 & 380 & 1 38  1 38  1 38  1 38  1 38  1 38  1 38  1 38  1 38  1 38  4 76  4 76  4 76  4 76  4 76  
\\ & &4 76  4 76  4 76  4 76  4 76  2 95  2 95  2 95  2 95  2 95  2 95  2 95  2 95  2 95  2 95  \\
31 & 380 & 1 38  1 38  1 38  1 38  1 38  1 38  1 38  1 38  1 38  1 76  1 76  2 76  2 76  2 76  3 76  
\\ & &     3 76  4 76  4 76  4 76  8 76  2 95  2 95  2 95  2 95  2 95  2 95  2 95  4 95  4 95  4 95  
\\ & &     4 95  \\
32 & 380 & 1 38  1 38  1 38  1 38  1 38  1 38  1 38  1 38  2 38  2 38  1 76  1 76  2 76  2 76  2 76  
\\ & &     2 76  2 76  2 76  2 76  2 76  2 76  1 95  3 95  4 95  4 95  4 95  4 95  4 95  4 95  4 95
\\ & &       4 95  4 95  \\
33 & 1824& 2 38  1 48  1 48  1 48  1 48  1 48  2 48  2 48  2 48  2 48  2 48  1 57  1 57  1 57  1 57  
\\ & &     1 57  1 57  2 57  2 57  2 57  2 57  2 57  3 96  3 96  3 96  3 96  3 96  3 96  3 96  3 96
\\ & &       3 96  3 96  4 96  \\
34 & 380 & 1 38  1 38  1 38  1 38  1 38  1 38  1 38  1 38  1 38  1 38  1 76  1 76  2 76  2 76  2 76  
\\ & &     2 76  2 76  2 76  2 76  2 76  3 76  3 76  2 95  2 95  2 95  2 95  4 95  4 95  4 95  4 95
\\ & &       4 95  4 95  4 95  4 95  \\
35 & 2400& 1 48  1 48  1 48  1 48  1 48  1 48  1 48  1 48  1 48  2 48  2 48  2 48  2 75  2 75  2 75  
\\ & &     2 75  2 75  2 75  2 75  2 75  3 75  3 75  3 75  1 96  1 96  1 96  3 96  3 96  3 96  3 96  
\\ & &     3 96  3 96  3 96  3 96  7 96  \\
36 & 2400& 1 48  1 48  1 48  1 48  1 48  1 48  1 48  2 48  2 48  2 48  2 48  2 48  1 75  1 75  1 75  
\\ & &     1 75  1 75  2 75  3 75  3 75  3 75  3 75  3 75  3 75  2 96  2 96  2 96  2 96  2 96  2 96  
\\ & &     3 96  3 96  3 96  3 96  3 96  3 96  \\
37 & 2400& 1 48  1 48  1 48  1 48  1 48  1 48  1 48  1 48  1 48  1 48  3 48  3 48  1 75  1 75  2 75  
\\ & &     2 75  2 75  2 75  2 75  2 75  2 75  2 75  3 75  4 75  1 96  2 96  2 96  2 96  2 96  2 96  
\\ & &     3 96  3 96  3 96  3 96  3 96  3 96  3 96 \\
38 & 2400& 1 48  1 48  1 48  1 48  1 48  1 48  1 48  1 48  1 48  1 48  2 48  2 48  1 75  1 75  2 75  
\\ & &     2 75  2 75  2 75  2 75  2 75  2 75  2 75  2 75  2 75  3 75  1 96  2 96  3 96  3 96  3 96  
\\ & &     3 96  3 96  3 96  3 96  3 96  3 96  3 96  3 96  \\
39 & 8400& 1 25  1 75  1 75  1 75  1 75  1 75  1 75  1 75  3 75  3 75  3 75  3 75  3 75  2 84  2 84  
\\ & &     2 84  2 84  2 84  2 84  2 84  2 84  2 84  2 84  2 84  3 84  3 84  1 96  1 96  1 96  1 96  
\\ & &     1 96  2 96  3 96  3 96  3 96  3 96  3 96  3 96  7 96  \\
\hline
\end{tabular}

  \caption{Solutions}
  \label{tab:solutions}
\end{table}

The computations described in this note are performed using the
finite-domain constraint compiler \bee~\cite{Metodijair2013} which compiles
constraints to a CNF, and solves it applying an underlying SAT
solver. We use Glucose 4.0~\cite{Glucose}.
All computations were performed on an Intel E8400 core, clocked at 2
GHz, able to run a total of $12$ parallel threads. Each of the cores
in the cluster has computational power comparable to a core on a
standard desktop computer.  Each SAT instance is run on a single
thread, and all running times reported in this paper are CPU times.

Table~\ref{tab:bee} describes the experimental evaluation. The first
two columns describe the instance: $n$ and the maximum value of a
common multiple in the solution. The column titled ``\bee'' is the
compile time (seconds) from constraints to CNF. The next two columns
specify the CNF size in number of clauses and variables. The right
most column specifies the SAT solving time in seconds (except where
marked as hours).

In the experiments we search for suitable values of $maxL$. Basically,
for smaller values of $n$, we start from 100 and increment by 100
until a solution is found. For larger values of $n$, we start from
1000 and increment by 500, and then refine the value from the largest
multiple of 1000 that has a solution incrementing by 100.

Table~\ref{tab:solutions} details the solutions found using our
encoding. The first column details the number $n$ of fractions. The
second column details the common multiplier (the value of $L$) in the
solution found. The third column details the solution found.
Note that for $n<3$ there is no solution as the constraint that states
that the number of occurrences of each digit in $\{1,\ldots,9\}$ is
between 1 and $\lceil n/3\rceil$ is trivially violated.

%\bibliographystyle{splncs}
%\bibliography{paper}

\end{document}